\documentclass[twocolumn,secnumarabic,amssymb, nobibnotes, aps, prb]{revtex4-1}
\usepackage{amsmath}
\usepackage{graphicx}
\usepackage[colorlinks=true,linktocpage=true,breaklinks=true]{hyperref}
\usepackage{xcolor}
\definecolor{darkgreen}{rgb}{0.0, 0.5, 0.0}

\begin{document}

\title{A new decomposition of the Kubo-Bastin formula}%

\author{Varga Bonbien$^1$}%
\email[]{bonbien.varga@kaust.edu.sa}
\author{Aur\'elien Manchon$^{1,2}$}
\email[]{manchon@cinam.univ-mrs.fr}
\affiliation{$^1$Physical Science and Engineering Division (PSE), King Abdullah University of Science and Technology (KAUST), Thuwal 23955-6900, Saudi Arabia\\
$^2$Aix-Marseille Univ, CNRS, CINaM, Marseille, France}
\begin{abstract}
The Smrcka-Streda version of Kubo's linear response formula is widely used in the literature to compute non-equilibrium transport properties of heterostructures. It is particularly useful for the evaluation of intrinsic transport properties associated with the Berry curvature of the Bloch states, such as anomalous and spin Hall currents as well as the damping-like component of the spin-orbit torque. Here, we demonstrate in a very general way that the widely used decomposition of the Kubo-Bastin formula introduced by Smrcka and Streda contains an overlap, which has lead to widespread confusion in the literature regarding the Fermi surface and Fermi sea contributions. To remedy this pathology, we propose a new decomposition of the Kubo-Bastin formula based on the permutation properties of the correlation function and derive a new set of formulas, without an overlap, that provides direct access to the transport effects of interest. We apply these new formulas to selected cases and demonstrate that the Fermi sea and Fermi surface contributions can be uniquely addressed with our symmetrized approach.
\end{abstract}
\maketitle

\section{Introduction}
The seminal work of Kubo \cite{Kubo1956,Kubo1957} showed that, in the perturbative weak-field limit, transport coefficients can be expressed as correlation functions of quantum mechanical observable operators. The resulting Kubo formalism has become a staple of quantum transport theory calculations and surged in popularity following the realization that applying it to transport phenomena in crystals provides direct access to topological invariants, thereby yielding an explanation for the robustness of the quantized Hall effect \cite{TKNN}.

While the original Kubo formula is formally satisfying, realistic calculations with it are rather impractical. \citet{Bastin1971} and later Streda and Smrcka \cite{SmrckaStreda1975} used Green's functions to rewrite the Kubo formula and arrived at a result directly applicable to computations in the static limit. Later on, Smrcka and Streda \cite{SmrckaStreda1977} further decomposed the Bastin formula into two terms, 
\begin{eqnarray}\label{eq:ssintroI}
{\cal A}_I&=&\frac{\hbar}{2\pi}\int d\varepsilon \partial_\varepsilon f(\varepsilon) {\rm Re}\left\{{\rm tr}[\hat{A}\hat{G}^r\hat{B}(\hat{G}^r-\hat{G}^a)]\right\},\\
{\cal A}_{II}&=&\frac{\hbar}{2\pi}\int d\varepsilon f(\varepsilon){\rm Re}\left\{{\rm tr}[\hat{A}\hat{G}^r\hat{B}\partial_\varepsilon\hat{G}^r-\hat{A}\partial_\varepsilon\hat{G}^r\hat{B}\hat{G}^r]\right\},\nonumber\\\label{eq:ssintroII}
\end{eqnarray}

Here ${\hat A}$ is the operator of the perturbation and ${\hat B}$ is the operator of the observable, $\hat{G}^{r(a)}(\varepsilon)$ is the retarded (advanced) Green's function of the system and we have suppressed the energy argument in the formulae for brevity, $f(\varepsilon)$ is the Fermi-Dirac distribution and $\partial_\varepsilon$ indicates an energy derivative. Because of their connection to $\partial_\varepsilon f(\varepsilon)$ and $f(\varepsilon)$, Eqs. \eqref{eq:ssintroI} and \eqref{eq:ssintroII} were wrongly referred to as {\em Fermi surface} and {\em Fermi sea} terms, respectively. These terms were used by Streda in his famous analysis of the quantized Hall effect \cite{Streda1982}. More recently, Cr\'epieux and Bruno \cite{CrepieuxBruno2001} presented a detailed and widely cited derivation of the Kubo-Bastin and Smrcka-Streda formulae from the Kubo formula.

The Smrcka-Streda formula has been widely used to compute charge and spin Hall currents \cite{NagaosaAHErev2010,Sinova2015} as well as spin-orbit torques \cite{Manchon2019}. Whereas a few works use the full Smrcka-Streda formula \cite{Freimuth2014a,Ghosh2018,Manchon2020,Geranton2015}, most theoretical studies exploit a simplified version of it, obtained by assuming constant scattering time and in the weak disorder limit \cite{Zelezny2014,Zelezny2017b,Li2015b}
\begin{eqnarray}\label{eq:simpleI}
{\cal A}_{surf}^{\Gamma}&\rightarrow&\frac{1}{\pi}\sum_{{\bf k}, n,m}\frac{\Gamma^2 {\rm Re}[\langle n{\bf k}|\hat{B}|m{\bf k}\rangle\langle m{\bf k}|\hat{A}|n{\bf k}\rangle]}{[(\varepsilon_{\rm F}-\varepsilon_{n{\bf k}})^2+\Gamma^2][(\varepsilon_{\rm F}-\varepsilon_{m{\bf k}})^2+\Gamma^2]},\\
{\cal A}_{sea}^{\Gamma}&\rightarrow&\sum_{{\bf k}, n\neq m}\frac{{\rm Im}[\langle n{\bf k}|\hat{B}|m{\bf k}\rangle\langle m{\bf k}|\hat{A}|n{\bf k}\rangle]}{(\varepsilon_{n{\bf k}}-\varepsilon_{m{\bf k}})^2}(f(\varepsilon_{n{\bf k}})-f(\varepsilon_{m{\bf k}})).\nonumber\\\label{eq:simpleII}
\end{eqnarray}
Here $\Gamma$ is the homogeneous broadening and $|n{\bf k}\rangle$ is a Bloch state of the crystal. This simplified version readily attributes ${\cal A}_{surf}^\Gamma$ to intraband transitions, yielding a $\sim 1/\Gamma$ dependence, and ${\cal A}_{sea}^\Gamma$ to interband transitions, which are finite in the clean limit ($\Gamma\rightarrow0$). In fact, these simplified formulae elegantly connect the Fermi sea transport contributions to the Berry curvature of the Bloch states, and, to date, the Berry curvature formula, Eq. \eqref{eq:simpleII}, has been widely used to characterize the intrinsic spin Hall effect of bulk materials \cite{Guo2008,Sahin2015,Sun2016}. As we mentioned already, this formula is only valid in the clean limit and does not apply in realistic materials where momentum scattering is important. More specifically, it becomes invalid when the broadening $\Gamma$ is comparable to, or larger than the local orbital gaps resulting from avoided band crossings, and where Berry curvature is maximized. Indeed, further investigations\cite{Tanaka2008, Kontani2009, Kontani2007} have addressed the spin Hall effect of metals using the full Smrcka-Streda formula, Eqs. \eqref{eq:ssintroI}-\eqref{eq:ssintroII}, showing evidence that the spin Hall conductivity of 5d transition metals is dominated by ${\cal A}_{I}$ \cite{Kontani2007}. Similarly, an influential work by \citet{Sinitsyn2006} demonstrated that in the case of a gapped Dirac cone, spin Hall effect is entirely due to ${\cal A}_I$ in the metallic regime, while it is entirely due to ${\cal A}_{II}$ in the gap. These observations, valid for specific examples, led to the confusion that ${\cal A}_I$ always dominate in metals. For instance, some investigations \cite{Freimuth2014a, Ghosh2018, Manchon2020} (including ours) have computed the spin-orbit torque using only ${\cal A}_I$ based on \citet{Kontani2007}'s argument. However, recent calculations have demonstrated that certain transport properties associated with Berry curvature, such as the dampinglike torque in magnetic heterostructures, have contributions from both ${\cal A}_I$ and ${\cal A}_{II}$\cite{Ghosh2019, Mahfouzi2018a, Wimmer2016} (see also Ref. \onlinecite{Turek2014}). This suggests that attributing purely Fermi surface origin to ${\cal A}_I$ and purely Fermi sea origin to ${\cal A}_{II}$ is incorrect.\\ 

In this paper, we first show in a very general way that the ${\cal A}_I$-${\cal A}_{II}$ decomposition of the Kubo-Bastin formula introduced by Smrcka and Streda contains an overlap, and there appears to be widespread confusion regarding this aspect in the literature. This overlap was hinted at for the special case of a 2-dimensional Dirac material by \citet{Sinitsyn2007}, but the fact that Smrcka-Streda and many subsequent authors unjustifiably neglected a subtle term relating to position operators in certain versions of the Smrcka-Streda formula responsible for geometric effects, went unmentioned \cite{Streda1982,CrepieuxBruno2001,Turek2012}. This subtlety is unnoticeable for simple models --- such as the quadratic magnetic Rashba gas --- when ${\cal A}_{II}$ is vanishingly small away from the avoided band crossing, since the neglected geometric term exactly cancels out Streda's orbital sea term \cite{Streda1982} in ${\cal A}_{II}$ which, due to the overlap, also appears in the ${\cal A}_I$ term. However, in the general case, ${\cal A}_{II}$ is non-negligible \cite{Turek2014,Kodderitzsch2015} and thus, Smrcka and Streda's decomposition of the Kubo-Bastin formula into ${\cal A}_I,{\cal A}_{II}$ does not lend itself to a proper analysis of different physical effects. To remedy this, we propose a new decomposition of the Kubo-Bastin formula based on the permutation properties of the correlator and derive a new set of formulas without an overlap, that provides direct access to the intrinsic geometric effects.

\section{The Kubo-Bastin formula and the Smrcka-Streda decomposition}

The Kubo-Bastin formula for the electrical conductivity, $\sigma_{kl}$, in the static limit as obtained from the Kubo formula is [Eq. (A9) of Ref. \onlinecite{CrepieuxBruno2001}]

\begin{eqnarray}
\label{eq:Bastin}
\sigma_{kl} = -\frac{\hbar}{2\pi}\int d\varepsilon\,f(\varepsilon)\,\text{tr}\bigg(&\big({\hat j}_k\partial_{\varepsilon}\hat{G}^{r}{\hat j}_l-{\hat j}_l\partial_{\varepsilon}\hat{G}^{a}{\hat j}_k\big)
\nonumber\\
&\times\big(\hat{G}^r-\hat{G}^a\big)\bigg),
\end{eqnarray}

where ${\hat j}_k,{\hat j}_l$ are electric charge current operators in the $k,l\in\{x,y,z\}$ directions, $\hat{G}^{r(a)}(\varepsilon) = \lim_{\eta\to 0}1/(\varepsilon-\mathcal{H}_0\pm i\eta)$ is the retarded(advanced) Green's function corresponding to the equilibrium Hamiltonian $\mathcal{H}_0$ and $\partial_\varepsilon\hat{G}^{r(a)}(\varepsilon)$ is the derivative of the Green's function with respect to its energy argument, that we have suppressed in the formula for brevity. Across the manuscript, $\hat{...}$ denotes an operator and {\rm tr}$(\dots)$ is the trace operation.\\

Splitting (\ref{eq:Bastin}) into two halves, integrating one of them by parts and combining it with the other half yields the Smrcka-Streda decomposition of the Kubo-Bastin formula [Eq. (A10) of Ref. \onlinecite{CrepieuxBruno2001}] with $\sigma_{kl}=\sigma_{kl}^{I}+\sigma_{kl}^{II}$, where

\begin{equation}
\label{eq:KuboStredaSurface}
\begin{split}
\sigma_{kl}^{I}&=\frac{\hbar}{4\pi}\int d\varepsilon\,\partial_{\varepsilon}f(\varepsilon)\,\text{tr}\bigg(\big({\hat j}_k\hat{G}^{r}{\hat j}_l-{\hat j}_l\hat{G}^{a}{\hat j}_k\big)
\\
&\qquad\qquad\qquad\qquad\qquad\qquad\times\big(\hat{G}^r-\hat{G}^a\big)\bigg)
\end{split}
\end{equation}

and

\begin{equation}
\label{eq:KuboStredaSea}
\begin{split}
\sigma_{kl}^{II}&=\frac{\hbar}{4\pi}\int d\varepsilon\,f(\varepsilon)\,\text{tr}\bigg({\hat j}_k\hat{G}^{r}(\varepsilon){\hat j}_l\partial_{\varepsilon}\hat{G}^{r}-{\hat j}_k\partial_{\varepsilon}\hat{G}^{r}{\hat j}_l\hat{G}^{r}
\\
&\qquad\qquad+{\hat j}_l\partial_{\varepsilon}\hat{G}^{a}{\hat j}_k\hat{G}^{a}-{\hat j}_l\hat{G}^{a}{\hat j}_k\partial_{\varepsilon}\hat{G}^{a}\bigg).
\end{split}
\end{equation}

Integrating (\ref{eq:KuboStredaSea}) by parts shall not yield any surface terms, thus we might naively conclude that this term describes effects resulting purely from the sea. However, this is not the case, since there is significant overlap between $\sigma_{kl}^{I}$ and $\sigma_{kl}^{II}$. Indeed, manipulating (\ref{eq:KuboStredaSurface}) and (\ref{eq:KuboStredaSea}) we arrive at (see Appendix)

\begin{subequations}
\begin{align}
\label{eq:Surf}
&\sigma_{kl}^{I}=\sigma_{kl}^{surf}+\sigma_{kl}^{ol},
\\
\label{eq:Sea}
&\sigma_{kl}^{II}=\sigma_{kl}^{sea}-\sigma_{kl}^{ol},
\end{align}
\end{subequations}

where

\begin{eqnarray}
\label{eq:KuboStredaSymm}
\sigma_{kl}^{surf}=\frac{\hbar}{4\pi}\int d\varepsilon\,\partial_{\varepsilon}f(\varepsilon)\,\text{tr}\bigg(&{\hat j}_k\big(\hat{G}^{r}-\hat{G}^{a}\big){\hat j}_l
\nonumber\\
&\times\big(\hat{G}^r-\hat{G}^a\big)\bigg),
\end{eqnarray}

\begin{eqnarray}
\label{eq:KuboStredaAsymm}
\sigma_{kl}^{sea}=-\frac{\hbar}{4\pi}\int d\varepsilon\,f(\varepsilon)\,&\text{tr}\bigg(\bigg\{{\hat j}_k\big(\partial_{\varepsilon}\hat{G}^{r}+\partial_{\varepsilon}\hat{G}^{a}\big){\hat j}_l
\nonumber\\
&-{\hat j}_l(\partial_{\varepsilon}\hat{G}^{r}+\partial_{\varepsilon}\hat{G}^{a}\big){\hat j}_k\bigg\}
\nonumber\\
&\times\big(\hat{G}^r-\hat{G}^a\big)\bigg)
\end{eqnarray}

and the overlap term

\begin{eqnarray}
\label{eq:KuboStredaOverlap}
\sigma_{kl}^{ol}=\frac{\hbar}{8\pi}\int d\varepsilon\,\partial_{\varepsilon}f(\varepsilon)\,&\text{tr}\bigg(\bigg\{{\hat j}_k\big(\hat{G}^{r}+\hat{G}^{a}\big){\hat j}_l
\nonumber\\
&-{\hat j}_l(\hat{G}^{r}+\hat{G}^{a}\big){\hat j}_k\bigg\}
\nonumber\\
&\times\big(\hat{G}^r-\hat{G}^a\big)\bigg).
\end{eqnarray}

Upon closer inspection, we note that $\sigma_{kl}^{surf}$ is symmetric whereas $\sigma_{kl}^{sea}$ along with $\sigma_{kl}^{ol}$ are antisymmetric under the exchange of operators ${\hat j}_k$ and ${\hat j}_l$. Furthermore, in the special case of $k=l$, $\sigma_{kk}^{surf}$ can be recognized as the Kubo-Greenwood formula for the diagonal conductivity. The separation of $\sigma^I$ into symmetric and antisymmetric parts yielding $\sigma^{surf}$ and $\sigma^{ol}$ is already present in the literature \cite{Fujimoto2014,Kodderitzsch2015}, however it was not realized that the antisymmetric part $\sigma^{ol}$ is an overlap and gets exactly cancelled when considering an appropriate separation of $\sigma^{II}$ into $\sigma^{sea}$ and $\sigma^{ol}$, as considered here.\\

In order to gain some understanding of $\sigma_{kl}^{sea}$ and $\sigma_{kl}^{ol}$ we use the expressions ${\hat j}_k=-ie/\hbar[\hat{G}^{-1},\hat{x}_k]$, where $\hat{x}_k$ is the position operator and $\partial_{\varepsilon}\hat{G}^{r(a)}=-(\hat{G}^{r(a)})^2$. In the clean limit ($\hat{G}^r\hat{G}^{-1}\to \hat{1},\,\hat{G}^a\hat{G}^{-1}\to\hat{1}$), the overlap term $\sigma_{kl}^{ol}$ from (\ref{eq:KuboStredaOverlap}) becomes

\begin{equation}
\label{eq:KuboStredaOverlapSimp}
\sigma_{kl}^{ol}\to\frac{ie}{4\pi}\int d\varepsilon\,\partial_{\varepsilon}f(\varepsilon)\,\text{tr}\bigg(\big(\hat{G}^r-\hat{G}^a\big)\big(\hat{x}_k{\hat j}_l-\hat{x}_l{\hat j}_k\big)\bigg),
\end{equation}

whereas $\sigma_{kl}^{sea}$ from (\ref{eq:KuboStredaAsymm}) simplifies to

\begin{equation}
\label{eq:KuboStredaAsymmSimp}
\sigma_{kl}^{sea}\to-\frac{e^2}{2\pi\hbar}\int d\varepsilon\,f(\varepsilon)\,\text{tr}\bigg(\big(\hat{G}^r-\hat{G}^a\big)[\hat{x}_k,\hat{x}_l]\bigg).
\end{equation}

We recognize (\ref{eq:KuboStredaOverlapSimp}) as Streda's orbital sea term \cite{Streda1982}. However, contrary to the original derivation in Ref. \onlinecite{SmrckaStreda1977} as well as in the re-derivation in Ref. \onlinecite{CrepieuxBruno2001} (see also Refs. \onlinecite{Turek2012}, \onlinecite{Kodderitzsch2013}, \onlinecite{Chadova2017}), this is not equivalent to $\sigma_{kl}^{II}$ but, as seen in Eq. \eqref{eq:Sea}, is an overlap term which has no overall effect since it gets cancelled out. Indeed, looking at Appendix A of Ref. \onlinecite{CrepieuxBruno2001}, we see that their $\tilde{\sigma}^I$ from Eq. (A11) is the same as our $\sigma^{I}$ in Eq. \eqref{eq:KuboStredaSurface}, but their $\tilde{\sigma}^{II}$ in Eq. (A12), which should be the total $\sigma^{II}$ is only our overlap term $-\sigma^{ol}$. In other words, something was 'lost'  while going from the general term $\sigma^{II}$ --- the second integral in their Eq. (A10)  and our Eq. (\ref{eq:KuboStredaSea}) --- to the 'simplified' or orbital sea term that is their Eq. (A12) and what we call the 'overlap' term in Eq. \eqref{eq:KuboStredaOverlapSimp}. What was 'lost' is precisely $\sigma^{sea}$, expressed in Eq. \eqref{eq:KuboStredaAsymm} and in the clean limit as Eq. \eqref{eq:KuboStredaAsymmSimp}, due to the fact that the position operators were assumed to commute. However, the latter is not necessarily true, since the weighting with the Fermi-Dirac distribution projects the total space of states to the filled states, and such terms containing non-commuting position operators are responsible for certain geometric effects such as those stemming from the Berry curvature \cite{XiaoBerryRev2010,Parker2019}.

Then why is it that, even though Streda's orbital sea term -- what we call the 'overlap' term -- from (\ref{eq:KuboStredaOverlapSimp}) has no overall effect and the geometric term (\ref{eq:KuboStredaAsymmSimp}) has been neglected in the literature, it is still possible to obtain proper results including Berry curvature effects for certain cases? In order to answer this question, consider the case of a vanishing $\sigma^{II}$ term: $\sigma_{kl}^{II}=0$, such as for the 2D metallic Dirac gas, or quadratic magnetic Rashba gas \cite{NagaosaAHErev2010,Papa2017}. From (\ref{eq:Surf}) and (\ref{eq:Sea}) we have $\sigma_{kl}^{II}=\sigma_{kl}^{sea}-\sigma_{kl}^{ol}=0\Rightarrow \sigma_{kl}^{sea}=\sigma_{kl}^{ol}$ giving $\sigma_{kl}^{I}=\sigma_{kl}^{surf}+\sigma_{kl}^{ol}=\sigma_{kl}^{surf}+\sigma_{kl}^{sea}$. Thus we see that for the particular case of a vanishing $\sigma^{II}$ term, Streda's orbital sea term (\ref{eq:KuboStredaOverlapSimp}) is exactly equal to the geometric term $\sigma_{kl}^{sea}$ and consequently describes Berry curvature effects. This is an advantage in the zero temperature case, since $\partial_{\varepsilon}f(\varepsilon)\to-\delta(\varepsilon-\varepsilon_F)$ as $T\to 0$, meaning that we can simply evaluate the Green's functions in (\ref{eq:KuboStredaOverlapSimp}) at the Fermi energy and there is no need for a complete energy integration, as would be required for (\ref{eq:KuboStredaAsymm}) or (\ref{eq:KuboStredaAsymmSimp}).\\

\section{The permutation decomposition}

Once we exclude pathological toy models from our investigations, such as the quadratic Rashba gas mentioned above, and turn our focus to real materials, the general sea term $\sigma_{kl}^{II}$ is strictly non-vanishing \cite{Turek2014,Kodderitzsch2015}, and so we propose not to consider the conventional Smrcka-Streda decomposition $\sigma_{kl}=\sigma_{kl}^{I}+\sigma_{kl}^{II}=(\sigma_{kl}^{surf}+\sigma_{kl}^{ol})+(\sigma_{kl}^{sea}-\sigma_{kl}^{ol})$ with the overlap term in any capacity. Rather, we offer a new one, the permutation decomposition:

\begin{equation}
\sigma_{kl}=\sigma_{kl}^{surf}+\sigma_{kl}^{sea},
\end{equation}

where $\sigma_{kl}^{surf}$ and  $\sigma_{kl}^{sea}$ are expressed in Eqs. (\ref{eq:KuboStredaSymm}) and (\ref{eq:KuboStredaAsymm}) respectively. As briefly mentioned above, $\sigma_{kl}^{surf}$ is symmetric whereas $\sigma_{kl}^{sea}$ is antisymmetric under the exchange of ${\hat j}_k$ and ${\hat j}_l$. Due to $\sigma_{kl}^{surf}$ and $\sigma_{kl}^{sea}$ being in different permutation classes they cannot overlap, and so they can be derived directly from the Bastin formula in Eq. (\ref{eq:Bastin}) by decomposing the latter into symmetric and antisymmetric terms with respect to the permutation of ${\hat j}_k$ and ${\hat j}_l$, effectively foregoing the need to go through the Smrcka-Streda decomposition and all subsequent analysis.\\

To see the direct derivation explicitly, we first symmetrize (\ref{eq:Bastin})

\begin{equation}
\sigma_{kl}=\frac{1}{2}(\sigma_{kl}+\sigma_{lk})+\frac12(\sigma_{kl}-\sigma_{lk}).
\end{equation}

It is important to add that although the notation suggests symmetrizing the cartesian indices of the conductivity tensor, we are in fact exchanging the operators themselves. In the given case, these are equivalent since the two current operators ${\hat j}_k,{\hat j}_l$ only differ in their direction. The distinction is, however, crucial for other cases, such as the spin response to an electric field, where the two operators under consideration are not the same, but are in fact $\hat{s}_k,{\hat j}_l$, where $\hat{s}_k$ is the spin operator in the $k$ direction, instead of ${\hat j}_k,{\hat j}_l$.\\

The symmetric part becomes

\begin{equation}
\label{eq:BastinSymm}
\begin{split}
\sigma_{kl}^{surf} = &-\frac{\hbar}{2\pi}\int d\varepsilon\,f(\varepsilon)\,\frac12\text{tr}\bigg(\bigg\{{\hat j}_k(\partial_{\varepsilon}\hat{G}^{r}-\partial_{\varepsilon}\hat{G}^{a}){\hat j}_l
\\
&\qquad+{\hat j}_l(\partial_{\varepsilon}\hat{G}^{r}-\partial_{\varepsilon}\hat{G}^{a}){\hat j}_k\bigg\}\big(\hat{G}^r-\hat{G}^a\big)\bigg).
\end{split}
\end{equation}

Next, we use the following identity

\begin{equation}
\begin{split}
&\int d\varepsilon\,f(\varepsilon)\,\text{tr}\bigg(\bigg\{{\hat j}_k(\partial_{\varepsilon}\hat{G}^{r}-\partial_{\varepsilon}\hat{G}^{a}){\hat j}_l
\\
&\qquad\qquad+{\hat j}_l(\partial_{\varepsilon}\hat{G}^{r}-\partial_{\varepsilon}\hat{G}^{a}){\hat j}_k\bigg\}\big(\hat{G}^r-\hat{G}^a\big)\bigg)
\\
&\qquad=\int d\varepsilon\,\partial_{\varepsilon}f(\varepsilon)\,\text{tr}\bigg({\hat j}_k\big(\hat{G}^{r}-\hat{G}^{a}\big){\hat j}_l
\big(\hat{G}^r-\hat{G}^a\big)\bigg),
\end{split}
\end{equation}

that can be shown straightforwardly via integration by parts and the cyclicity of the trace, leading directly to the expression of $\sigma_{kl}^{surf}$ in (\ref{eq:KuboStredaSymm}). The antisymmetric part (\ref{eq:KuboStredaAsymm}) is obtained directly from the antisymmetrization of (\ref{eq:Bastin}) without any intermediate steps.\\

The terms arrived at in this way carry a physical interpretation. Consider the clean limit ($\Gamma\to 0$). In this case, $\sigma_{kl}^{surf}$ vanishes as is seen by using ${\hat j}_k=-ie/\hbar[\hat{G}^{-1},\hat{x}_k]$ in (\ref{eq:KuboStredaSymm}) and so is purely extrinsic. On the other hand, $\sigma_{kl}^{sea}$ does not vanish, reduces to (\ref{eq:KuboStredaAsymmSimp}) and so is an intrinsic contribution. In the general case of a material with impurities the intrinsic contribution thus arises purely from $\sigma_{kl}^{sea}$, which can be very helpful when trying to extract information from experimental results by comparing them to numerical calculations performed using the permutation decomposition.

A further utility of decomposing the Kubo formula into permutation classes is the possibility of dealing with distinct physical effects arising as higher order responses in a straightforward manner. This has been completed this for second order response and is currently under preparation.

\section{Application to Hall effects, spin currents and spin-orbit torque}
In this section, we compute the transport properties of three illustrative systems using the two different decompositions of the Kubo-Bastin formula, the Smrcka-Streda decomposition,
\begin{eqnarray}
{\cal A}_I&=&\frac{\hbar}{2\pi}\int d\varepsilon \partial_\varepsilon  f(\varepsilon){\rm Re}\left\{{\rm tr}[\hat{A}\hat{G}^r\hat{B}(\hat{G}^r-\hat{G}^a)]\right\},\nonumber\\\label{eq:AI}\\
{\cal A}_{II}&=&\frac{\hbar}{2\pi}\int d\varepsilon f(\varepsilon){\rm Re}\left\{{\rm tr}[\hat{A}\hat{G}^r\hat{B}\partial_\varepsilon\hat{G}^r-\hat{A}\partial_\varepsilon\hat{G}^r\hat{B}\hat{G}^r]\right\},\nonumber\\\label{eq:AII}
\end{eqnarray}

and our new permutation decomposition,

\begin{eqnarray}
{\cal A}_{\rm surf}&=&\frac{\hbar}{4\pi}\int d\varepsilon \partial_\varepsilon  f(\varepsilon){\rm Re}\left\{{\rm tr}[\hat{A}(\hat{G}^r-\hat{G}^a)\hat{B}(\hat{G}^r-\hat{G}^a)]\right\},\nonumber\\\label{eq:Aa}\\
{\cal A}_{\rm sea}&=&\frac{\hbar}{2\pi}\int d\varepsilon f(\varepsilon){\rm Re}\left\{{\rm tr}[\hat{A}(\hat{G}^r-\hat{G}^a)\hat{B}(\partial_\varepsilon\hat{G}^r+\partial_\varepsilon\hat{G}^a)]\right\}.\nonumber\\\label{eq:Ab}
\end{eqnarray}

As discussed in the previous section, it is clear that ${\cal A}_I+{\cal A}_{II}={\cal A}_{\rm surf}+{\cal A}_{\rm sea}$. Now, we would like to show how the new separation can specifically distinguish between extrinsic and intrinsic phenomena. To do so, we consider non-equilibrium transport (i) in the magnetic Rashba gas, (ii) in a multiorbital tight-binding model of a ferromagnet/normal metal heterostructure and (iii) in a non-collinear antiferromagnet.

\subsection{Magnetic Rashba gas}
Let us first consider the canonical magnetic Rashba gas regularized on a square lattice and described by the Hamiltonian
\begin{eqnarray}
{\cal H}=&&-2t(\cos k_x+\cos k_y)+\Delta\hat{\sigma}_z\\
&&+t_{\rm R}(\hat{\sigma}_x\sin k_y-\hat{\sigma}_y\sin k_x).\nonumber
\end{eqnarray}
Here $t$ is the nearest-neighbor hopping, $t_{\rm R}$ is the Rashba parameter, and $\Delta$ is the s-d exchange. This model has been central to the investigation of the anomalous Hall effect \cite{Sinova2004,Onoda2008} and spin-orbit torque \cite{Manchon2008b,Qaiumzadeh2015}. Here, we do not consider the vertex correction since our interest is to illustrate the superiority of our new permutation decomposition of the Kubo-Bastin formula. The Green's function is simply given by $\hat{G}^{r(a)}(\varepsilon)=(\varepsilon-\hat{\cal H}\pm i\Gamma)^{-1}$, $\Gamma$ being the homogeneous broadening coming from short-range (delta-like) impurities. In this section, we compute the non-equilibrium properties induced by the electric field ($\hat{A}=-e \hat{j}_x$), with particular focus on the longitudinal conductivity ($\hat{B}=-e \hat{j}_x$), the transverse conductivity ($\hat{B}=-e \hat{j}_y$), the fieldlike torque ($\hat{B}=-\Delta\hat{\sigma}_y$) and the dampinglike torque ($\hat{B}=\Delta\hat{\sigma}_x$). The conductance of the two-dimensional electron gas is in $\Omega^{-1}$ and the spin torque is expressed in terms of an effective spin conductivity $(\hbar/2e)~\Omega^{-1}\cdot m^{-1}$. Finally, for the parameters we take $t_{\rm R}=2.4t$, $\Delta=0.2t$ and $\Gamma=0.1t$.\par

\begin{figure}
\begin{center}
        \includegraphics[width=9cm]{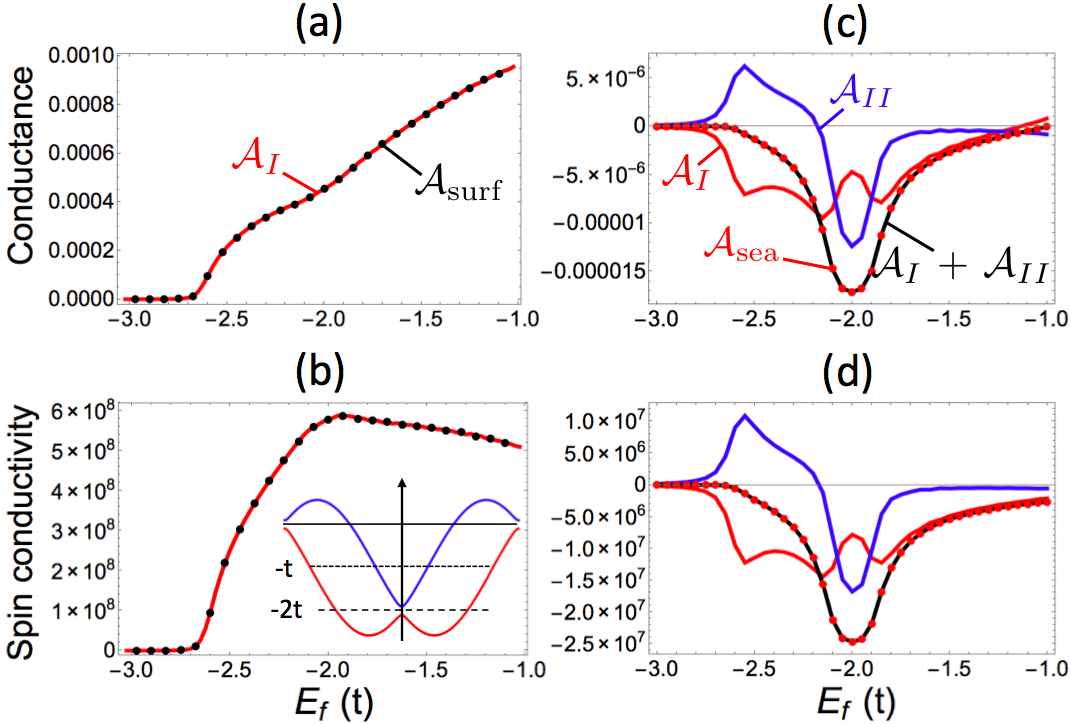}
      \caption{(Color online) Energy dependence of (a) longitudinal conductivity, (b) fieldlike torque, (c) transverse conductivity and (d) dampinglike torque in the two-dimensional magnetic Rashba gas. The solid red (blue) curve refers to the ${\cal A}_I$ (${\cal A}_{II}$) contribution, whereas the black curve is their sum ${\cal A}_I+{\cal A}_{II}$. The black (red) dots refer to ${\cal A}_{\rm surf}$ (${\cal A}_{\rm sea}$). The inset of (b) shows the band structure of the magnetic Rashba gas. The dashed horizontal line indicates the position of the avoided band crossing and the dotted line stands for the maximum energy taken in this calculation. The conductivity is in $\Omega^{-1}\cdot m^{-1}$ and the spin conductivity is in $(\hbar/2e)~\Omega^{-1}\cdot m^{-1}$. \label{Fig1}}
\end{center}
\end{figure}

Figure \ref{Fig1} reports the (a) longitudinal and (b) transverse Hall conductivities as well as the torque components, (b) fieldlike and (d) dampinglike, as a function of the energy. In this figure and the ones following, the ${\cal A}_I$ and ${\cal A}_{II}$ contributions of the Smrcka-Streda formula are represented with red and blue solid lines, while the Fermi surface (${\cal A}_{\rm surf}$) and Fermi sea (${\cal A}_{\rm sea}$) contributions of our permutation decomposition of the Kubo-Bastin formula are represented by black and red dots, respectively. The black line represents the sum ${\cal A}_I+{\cal A}_{II}$. In the case of transport properties only involving the Fermi surface, such as the longitudinal conductivity [Fig. \ref{Fig1}(a)] and the fieldlike torque [Fig. \ref{Fig1}(b)], ${\cal A}_{II}={\cal A}_{\rm sea}=0$ and ${\cal A}_{I}={\cal A}_{\rm surf}$, so using either the conventional Smrcka-Streda decomposition or our permutation decomposition is equivalent. 

The transport properties involving Fermi sea are more interesting to consider. Indeed, as discussed in the previous section, it clearly appears that when using the conventional Smrcka-Streda formula, both ${\cal A}_I$ (red) and ${\cal A}_{II}$ (blue) contributions are equally important. In fact, the variations of ${\cal A}_{II}$ can be readily correlated with the band structure displayed in the inset of Fig. \ref{Fig1}(b). The ${\cal A}_{II}$ curve exhibits two peaks, one close to the bottom of the lowest band, where the dispersion is quite flat (around -2.5$t$), and one when the Fermi level lies in the local gap corresponding to the avoided crossing of the two bands [dashed line in the inset of Fig. \ref{Fig1}(b)]. Away from this local gap, ${\cal A}_{II}$ vanishes. This is an important observation because it indicates that the overlap contribution of the Smrcka-Streda formula is peaked close to locally flat bands, irrespective whether it is geometrically trivial (around -2.5$t$) or non-trivial (around -2$t$). When summing ${\cal A}_{I}$ and ${\cal A}_{II}$, the complex structure of ${\cal A}_{II}$ close to the bottom of the lowest band compensates ${\cal A}_{I}$ exactly, so that the total contribution ${\cal A}_{I}+{\cal A}_{II}={\cal A}_{\rm sea}$ has a much simpler overall structure and is peaked only at the local (geometrically non-trivial) gap, which illustrates the Berry curvature origin of this contribution. This simple calculation points out the dramatic need to compute {\em both} ${\cal A}_{I}$ and ${\cal A}_{II}$ contributions to obtain correct Fermi sea contributions such as dampinglike torque and anomalous Hall effect, whereas ${\cal A}_{\rm sea}$ contains these contributions in itself. 

\subsection{Transition metal bilayer}
The previous calculation shows that the contribution of ${\cal A}_{II}$ becomes particularly crucial when crossing local {\em flat bands}. Nonetheless, one might argue that this sensitivity is due to the simplicity of the Rashba model that only involves two bands of opposite chirality. To generalize these results, we now move on to a more complex system, a metallic bilayer made of two transition metal slabs and modeled using a multiorbital tight-binding model within the Slater-Koster two-center approximation. This model has been discussed in detail in Refs. \onlinecite{Manchon2020,Hajr2020} and here we only summarize its main features. The structure consists of two adjacent transition metal layers with bcc crystal structure and equal lattice parameter. The 10 d-orbitals are included and the tight-binding parameters are extracted from Ref. \onlinecite{Papaconstantopoulos2015}. Importantly, we consider atomic (Russell-Saunders) spin-orbit coupling, so that bulk and interfacial spin-orbit coupled transport are modeled in a realistic manner.\par

\begin{figure}
\begin{center}
        \includegraphics[width=6cm]{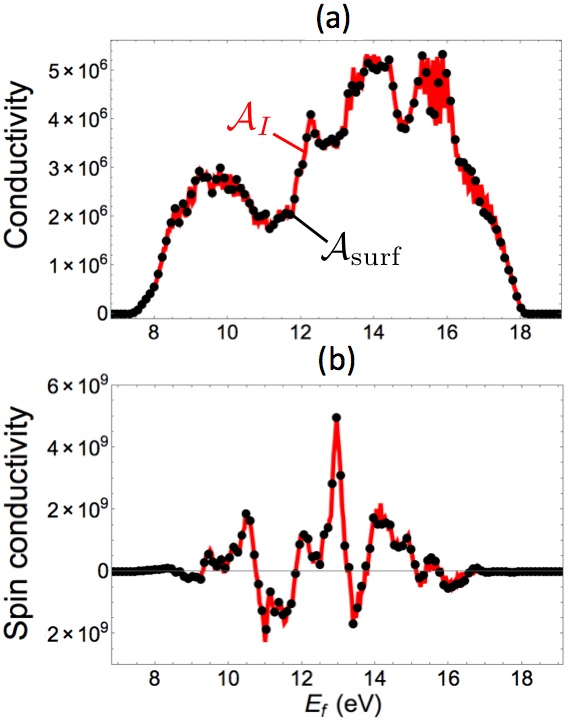}
      \caption{(Color online) Energy dependence of (a) longitudinal conductivity and (b) fieldlike torque in the multiorbital transition metal bilayer model. The solid red curve refers to the ${\cal A}_I$ and the black dots refer to ${\cal A}_{\rm surf}$. The conductivity is in $\Omega^{-1}\cdot m^{-1}$ and the spin conductivity is in $(\hbar/2e)~\Omega^{-1}\cdot m^{-1}$.\label{Fig2}}
\end{center}
\end{figure}

\begin{figure}
\begin{center}
        \includegraphics[width=9cm]{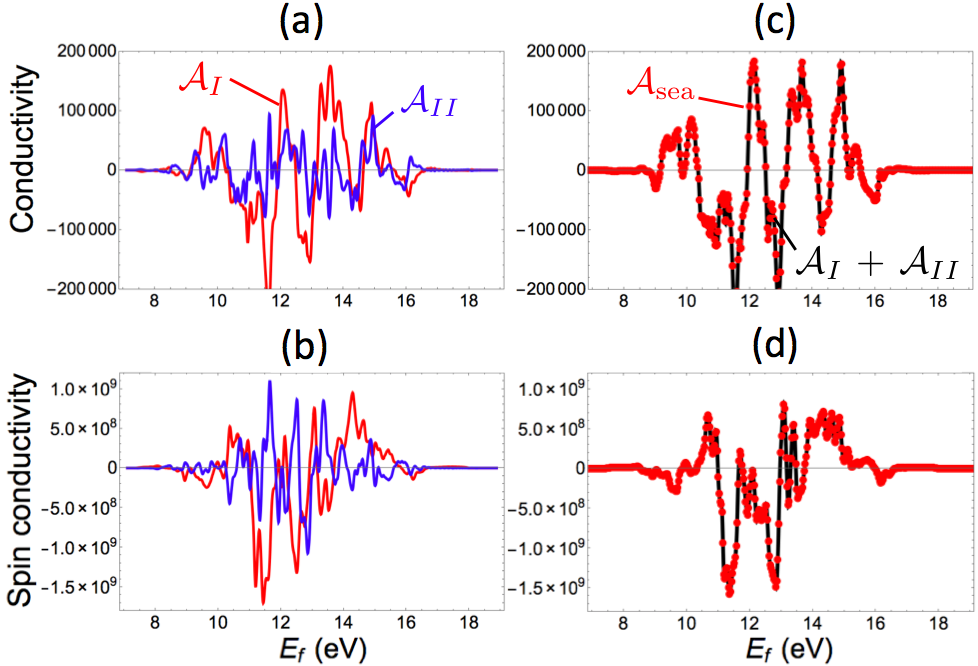}
      \caption{(Color online) Energy dependence of (a, b) transverse conductivity and (c, d) dampinglike torque in the multiorbital transition metal bilayer model. The solid red (blue) curve refers to the ${\cal A}_I$ (${\cal A}_{II}$) contribution, the black curve is their sum ${\cal A}_I+{\cal A}_{II}$ and the red dots refer to ${\cal A}_{\rm sea}$. The conductivity is in $\Omega^{-1}\cdot m^{-1}$ and the spin conductivity is in $(\hbar/2e)~\Omega^{-1}\cdot m^{-1}$.\label{Fig3}}
\end{center}
\end{figure}

Figure \ref{Fig2} reports the same transport properties as Fig. \ref{Fig1}, i.e., (a) longitudinal conductivity (i.e., the two-dimensional conductance divided by the thickness of the bilayer), as well as (b) the fieldlike torque as a function of the energy. Again, we find that Fermi surface properties are well-described by the surface terms when using either the conventional Smrcka-Streda or our permutation decomposition of the Kubo-Bastin formula [Fig. \ref{Fig2}(a, b)]. Nonetheless, the Fermi sea properties displayed on Fig. \ref{Fig3} exhibit a much richer behavior. The considerably more complex band structure of the multiorbital model (e.g., see Fig. 4 in Ref. \onlinecite{Manchon2020}) possesses a high density of flat band regions which results in highly oscillating ${\cal A}_{I}$ and ${\cal A}_{II}$ contributions, in both transverse conductivity [Fig. \ref{Fig3}(a)], and dampinglike torque [Fig. \ref{Fig3}(b)]. These oscillations are partially washed out when summing both contributions [Fig. \ref{Fig3}(c,d)] so that the remaining oscillations are only associated to the local Berry curvature of the band structure. These results agree with our recent work where we demonstrated, using a similar multi-band model for topological insulator/antiferromagnet heterostructures, that both ${\cal A}_{I}$ and ${\cal A}_{II}$ contributions are necessary to obtain the appropriate magnitude of the damping-torque, particularly in the regions displaying avoided band crossing \cite{Ghosh2019}. Figure \ref{Fig3} clearly shows that both contributions should be accounted for when computing dampinglike torque and anomalous transport. Taking only ${\cal A}_{I}$ into account like in Refs. \onlinecite{Freimuth2014a,Manchon2020} is insufficient. 

\subsection{Non-collinear antiferromagnet}

\begin{figure}
\begin{center}
        \includegraphics[width=6cm]{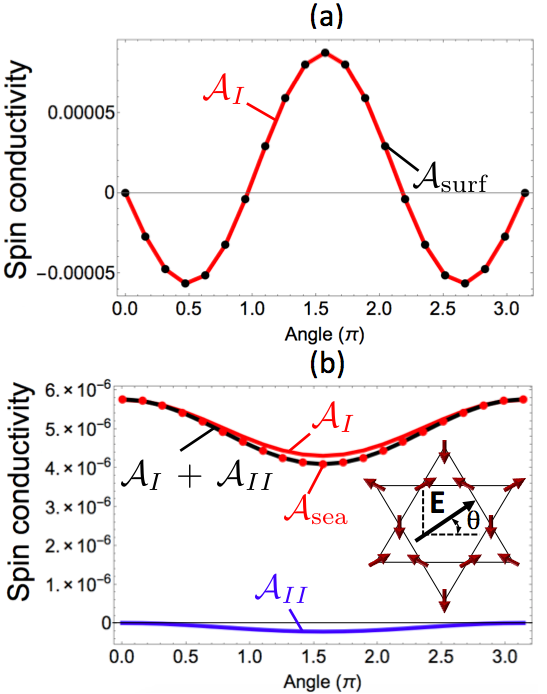}
      \caption{(Color online) Angular dependence of (a) in-plane and (b) out-of-plane spin Hall effect in the non-collinear antiferromagnetic Kagome lattice model. The solid red (blue) curve refers to the ${\cal A}_I$ (${\cal A}_{II}$) contribution, the black curve is their sum ${\cal A}_I+{\cal A}_{II}$ and the black (red) dots refer to ${\cal A}_{\rm surf}$ (${\cal A}_{\rm sea}$). The inset displays the angle made by the applied electric field with respect to the crystal axes. The spin conductivity is in $(\hbar/2e)~\Omega^{-1}$.\label{Fig4}}
\end{center}
\end{figure}

We conclude this investigation by considering one last system of interest: a non-collinear antiferromagnet displaying anomalous transverse spin currents even in the absence of spin-orbit coupling. As a matter of fact, the transport of spin and charge in non-collinear antiferromagnets has been the object of intense scrutiny recently, as anomalous Hall as well as magnetic spin Hall effects have been predicted\cite{Chen2014,Kubler2014,Zelezny2017b} and observed\cite{Nayak2015,Nakatsuji2015,Kimata2019}. We test our permutation decomposition on an ideal Kagome lattice with 120$^\circ$ magnetic moment configuration, as depicted in the inset of Fig. \ref{Fig4}. The model is the same as Ref. \onlinecite{Chen2014}, and the Hamiltonian reads
\begin{eqnarray}
{\cal H}&=&t\sum_{\langle i\alpha,j\beta\rangle}\hat{c}_{j\beta}^\dagger \hat{c}_{i\alpha}+\Delta \sum_i\hat{c}_{i\alpha}^\dagger\hat{\bm\sigma}\cdot{\bf m}_\alpha\hat{c}_{i\alpha}.
\end{eqnarray}
Here, $t$ is the nearest neighbor hopping, and $\Delta$ is the s-d exchange. The indices $\alpha,\beta$ refer to the different magnetic sublattices of a magnetic unit cell, and $i,j$ refer to different unit cells. In this work, we set $\Delta=1.7t$. Such a system displays two types of transverse spin currents\cite{Zelezny2017b,Zhang2018d}, even in the absence of spin-orbit coupling: one spin current $\sigma_s^z$ possesses a polarization perpendicular to the plane, and the other $\sigma_s^{\|}$ has a polarization in-plane and normal to the applied electric field. We refer to the former as perpendicular spin Hall current and the latter is called in-plane spin Hall current. 

We compute in Fig. \ref{Fig4} the (a) in-plane and (b) perpendicular spin conductivities as a function of the angle of the electric field with respect to the crystal lattice directions. We obtain that the in-plane spin current is purely a Fermi surface term, corresponding to the "magnetic spin Hall effect" predicted by \citet{Zelezny2017b} and observed by \citet{Kimata2019}. This spin current strongly depends on the orientation of the electric field with respect to the crystallographic axes. In contrast, the perpendicular spin current shows a weak angular dependence and is purely given by the Fermi sea contribution\cite{Zhang2018d}. Again, the ${\cal A}_{II}$ contribution is small but non-zero. The reduced magnitude of ${\cal A}_{II}$ compared to ${\cal A}_{I}$ is due to the fact that the Fermi level is taken away from the avoided band crossing in this particular case.

\section{Conclusion}
We have shown that the widely used Smrcka-Streda decomposition of the celebrated Kubo-Bastin formula possesses an overlap that makes it inappropriate to distinguish between Fermi sea and Fermi surface contributions to transport coefficients. This is particularly crucial in multiband systems possessing a high density of locally flat bands and avoided band crossings. As a matter of fact, whereas intrinsic (Berry-curvature induced) transport properties are dominated by geometrically non-trivial avoided band crossings, the overlap is enhanced close to any (trivial and non-trivial) locally flat bands, as illustrated in the case of the magnetic Rashba gas. Therefore, the Smrcka-Streda decomposition of the Kubo-Bastin formula can lead to an incorrect estimation of the intrinsic transport properties. To remedy this difficulty, we demonstrated that the Kubo formula can be decomposed into symmetric and antisymmetric parts, which gives direct access to Fermi surface and Fermi sea contributions. The superiority of this new permutation decomposition over Smrcka-Streda's, apart from its apparent conceptual clarity, has been illustrated by computing the extrinsic and intrinsic transport coefficients of three selected systems. This observation has substantial impact on quantum transport calculations, especially when considering Berry curvature induced mechanisms such as Hall conductance and torques, since it provides a neat way of separating the intrinsic part of these anomalous transport effects from Fermi surface related effects, removing spurious effects stemming from local trivial band flatness.

\begin{acknowledgements}
This work was supported by the King Abdullah University of Science and Technology (KAUST) through the award OSR-2017-CRG6-3390 from the Office of Sponsored Research (OSR). 
\end{acknowledgements}

\appendix*
\begin{widetext}
\section{Derivation of the overlap term}
The $\sigma^{I}$ term (\ref{eq:KuboStredaSurface}) can be handled in a simple way by separating it into symmetric and antisymmetric permutations of ${\hat j}_k$ and ${\hat j}_l$ as follows:

\begin{equation}
\label{eq:SurfOverlap1}
\begin{split}
\sigma_{kl}^{I}=&\frac{\hbar}{4\pi}\int d\varepsilon\,\partial_{\varepsilon}f(\varepsilon)\,\text{tr}\bigg(\big({\hat j}_k\hat{G}^{r}{\hat j}_l-{\hat j}_l\hat{G}^{a}{\hat j}_k\big)\big(\hat{G}^r-\hat{G}^a\big)\bigg)
\\
=&\frac{\hbar}{8\pi}\int d\varepsilon\,\partial_{\varepsilon}f(\varepsilon)\,\text{tr}\bigg(\big({\hat j}_k(\hat{G}^{r}-\hat{G}^{a}){\hat j}_l+{\hat j}_l(\hat{G}^{r}-\hat{G}^{a}){\hat j}_k\big)\big(\hat{G}^r-\hat{G}^a\big)\bigg)
\\
&+\frac{\hbar}{8\pi}\int d\varepsilon\,\partial_{\varepsilon}f(\varepsilon)\,\text{tr}\bigg(\big({\hat j}_k(\hat{G}^{r}+\hat{G}^{a}){\hat j}_l-{\hat j}_l(\hat{G}^{r}+\hat{G}^{a}){\hat j}_k\big)\big(\hat{G}^r-\hat{G}^a\big)\bigg)
\\
=
&\frac{\hbar}{4\pi}\int d\varepsilon\,\partial_{\varepsilon}f(\varepsilon)\,\text{tr}\bigg({\hat j}_k(\hat{G}^{r}-\hat{G}^{a}){\hat j}_l\big(\hat{G}^r-\hat{G}^a\big)\bigg)
\\
&+\frac{\hbar}{8\pi}\int d\varepsilon\,\partial_{\varepsilon}f(\varepsilon)\,\text{tr}\bigg(\big({\hat j}_k(\hat{G}^{r}+\hat{G}^{a}){\hat j}_l-{\hat j}_l(\hat{G}^{r}+\hat{G}^{a}){\hat j}_k\big)\big(\hat{G}^r-\hat{G}^a\big)\bigg)
\\
=&\sigma^{surf}_{kl}+\sigma^{ol}_{kl}.
\end{split}
\end{equation}

It is clear that $\sigma^{surf}$ is symmetric and $\sigma^{ol}$ is antisymmetric in the exchange of ${\hat j}_k$ and ${\hat j}_l$.\\
The $\sigma^{II}$ term (\ref{eq:KuboStredaSea}) is more complicated,  requiring the following manipulations:

\begin{equation}
\label{eq:SeaOverlap1}
\begin{split}
\sigma_{kl}^{II}&=\frac{\hbar}{4\pi}\int d\varepsilon\,f(\varepsilon)\,\text{tr}\bigg({\hat j}_k\hat{G}^{r}{\hat j}_l\partial_{\varepsilon}\hat{G}^{r}-{\hat j}_k\partial_{\varepsilon}\hat{G}^{r}{\hat j}_l\hat{G}^{r}
+{\hat j}_l\partial_{\varepsilon}\hat{G}^{a}{\hat j}_k\hat{G}^{a}-{\hat j}_l\hat{G}^{a}{\hat j}_k\partial_{\varepsilon}\hat{G}^{a}\bigg)
\\
&=\frac12\frac{\hbar}{4\pi}\int d\varepsilon\,f(\varepsilon)\,\text{tr}\bigg({\hat j}_k\hat{G}^{r}{\hat j}_l\partial_{\varepsilon}\hat{G}^{r}-{\hat j}_k\partial_{\varepsilon}\hat{G}^{r}{\hat j}_l\hat{G}^{r}
+{\hat j}_l\partial_{\varepsilon}\hat{G}^{a}{\hat j}_k\hat{G}^{a}-{\hat j}_l\hat{G}^{a}{\hat j}_k\partial_{\varepsilon}\hat{G}^{a}\bigg)
\\
&\quad+\frac12\frac{\hbar}{4\pi}\int d\varepsilon\,f(\varepsilon)\,\text{tr}\bigg({\hat j}_k\hat{G}^{r}{\hat j}_l\partial_{\varepsilon}\hat{G}^{r}-{\hat j}_k\partial_{\varepsilon}\hat{G}^{r}{\hat j}_l\hat{G}^{r}
+{\hat j}_l\partial_{\varepsilon}\hat{G}^{a}{\hat j}_k\hat{G}^{a}-{\hat j}_l\hat{G}^{a}{\hat j}_k\partial_{\varepsilon}\hat{G}^{a}\bigg)
\\
&=\frac12\frac{\hbar}{4\pi}\int d\varepsilon\,f(\varepsilon)\,\text{tr}\bigg({\hat j}_k(\hat{G}^{r}-\hat{G}^{a}){\hat j}_l(\partial_{\varepsilon}\hat{G}^{r}+\partial_{\varepsilon}\hat{G}^{a})-{\hat j}_k(\partial_{\varepsilon}\hat{G}^{r}+\partial_{\varepsilon}\hat{G}^{a}){\hat j}_l(\hat{G}^{r}-\hat{G}^{a})\bigg)
\\
&\quad+\frac12\frac{\hbar}{4\pi}\int d\varepsilon\,f(\varepsilon)\,\text{tr}\bigg({\hat j}_k(\hat{G}^{r}+\hat{G}^{a}){\hat j}_l(\partial_{\varepsilon}\hat{G}^{r}-\partial_{\varepsilon}\hat{G}^{a})-{\hat j}_k(\partial_{\varepsilon}\hat{G}^{r}-\partial_{\varepsilon}\hat{G}^{a}){\hat j}_l(\hat{G}^{r}+\hat{G}^{a})\bigg).
\end{split}
\end{equation}

Looking at the terms after the last equality in \eqref{eq:SeaOverlap1}, we integrate by parts the second term and combine the result with the first term. Some straightforward algebra yields

\begin{equation}
\begin{split}
\sigma_{kl}^{II}&=\frac{\hbar}{4\pi}\int d\varepsilon\,f(\varepsilon)\,\text{tr}\bigg({\hat j}_k(\hat{G}^{r}-\hat{G}^{a}){\hat j}_l(\partial_{\varepsilon}\hat{G}^{r}+\partial_{\varepsilon}\hat{G}^{a})-{\hat j}_k(\partial_{\varepsilon}\hat{G}^{r}+\partial_{\varepsilon}\hat{G}^{a}){\hat j}_l(\hat{G}^{r}-\hat{G}^{a})\bigg)
\\
&\quad-\frac12\frac{\hbar}{4\pi}\int d\varepsilon\,\partial_{\varepsilon}f(\varepsilon)\,\text{tr}\bigg({\hat j}_k(\hat{G}^{r}+\hat{G}^{a}){\hat j}_l(\hat{G}^{r}-\hat{G}^{a})-{\hat j}_k(\hat{G}^{r}-\hat{G}^{a}){\hat j}_l(\hat{G}^{r}+\hat{G}^{a})\bigg)
\\
&=-\frac{\hbar}{4\pi}\int d\varepsilon\,f(\varepsilon)\,\text{tr}\bigg(\big({\hat j}_k(\partial_{\varepsilon}\hat{G}^{r}+\partial_{\varepsilon}\hat{G}^{a}){\hat j}_l-{\hat j}_l(\partial_{\varepsilon}\hat{G}^{r}+\partial_{\varepsilon}\hat{G}^{a}){\hat j}_k\big)(\hat{G}^{r}-\hat{G}^{a})\bigg)
\\
&\quad-\frac{\hbar}{8\pi}\int d\varepsilon\,\partial_{\varepsilon}f(\varepsilon)\,\text{tr}\bigg(\big({\hat j}_k(\hat{G}^{r}+\hat{G}^{a}){\hat j}_l-{\hat j}_l(\hat{G}^{r}+\hat{G}^{a}){\hat j}_k\big)(\hat{G}^{r}-\hat{G}^{a})\bigg)
\\
&=\sigma^{sea}_{kl}-\sigma^{ol}_{kl}.
\end{split}
\end{equation}

\end{widetext}
\bibliography{Kspaper}

\end{document}